\documentclass[letterpaper,twocolumn,10pt]{article}
\usepackage{ligroup}
\newcommand{\mypara}[1]{\noindent\textbf{#1}}
\usepackage[hyphens]{url}
\usepackage{graphicx}
\usepackage{natbib}
\usepackage{caption}
\usepackage{booktabs}
\usepackage{tcolorbox}
\usepackage{amsmath,amssymb}
\usepackage{multirow}

\begin{document}

\date{}

\title{\bf Understanding Stage-Wise Utility-Risk Trade-offs \\in LLM Agent Memory}

\author{
Chuanchao Zang\textsuperscript{1}\ \ \
Zijian Cao\textsuperscript{1}\ \ \
Xiangtao Meng\textsuperscript{1}\ \ \
Jianing Wang\textsuperscript{1}\ \ \
Wenyu Chen\textsuperscript{1}\ \ \
\\
Xinyu Gao\textsuperscript{1}\ \ \
Li Wang\textsuperscript{1}\ \ \
Zheng Li\textsuperscript{1,2,3*}\ \ \
Shanqing Guo\textsuperscript{1,2,3*}\ \ \
\\
\\
\textsuperscript{1}\textit{School of Cyber Science and Technology, Shandong University}\\
\textsuperscript{2}\textit{State Key Laboratory of Cryptography and Digital Economy Security, Shandong University} \\
\textsuperscript{3}\textit{Shandong Key Laboratory of Artificial Intelligence Security, Shandong University}
}

\maketitle

\begin{abstract}
Long-term memory is becoming a core capability of LLM agents, enabling personalization and long-horizon interaction. However, memory mechanisms that retain, transform, or expose more information can affect both benign utility and susceptibility to memory poisoning. Existing evaluations typically measure memory utility or attack risk in isolation under fixed configurations, providing limited insight into how stage-specific design choices reshape their trade-off. We present \textsc{MemGauge}, a controllable framework that separately varies writing admission, management policy, and retrieval exposure under matched clean and poisoned conditions. Across 11 LLMs and two long-term memory benchmarks, controlled evaluations reveal three distinct profiles: a threshold-like risk transition during writing, policy-dependent local decoupling during management, and coupled growth of utility and risk during retrieval. We further apply analogous stage-level measurements to four existing memory systems and observe diagnostic associations qualitatively consistent with these profiles. These results show that targeted poisoning risk varies across memory operations and motivate stage-aware evaluation and control of LLM-agent memory.
\end{abstract}

\section{Introduction}
Long-term memory enables LLM agents to support personalization, accumulate experience, and interact over long horizons by retaining information beyond the current context\cite{park2023generative,zhong2024memorybank}.
Yet this persistence also creates security risks: once adversarially induced information enters memory, it can influence decisions across queries and sessions\cite{chen2024agentpoison,dong2026memory,dash2026untrusted}.
For example, Microsoft documented attempts to use hidden ``remember'' instructions to write promotional preferences into assistant memory, illustrating how a mechanism designed for personalization can also bias later recommendations \cite{kochavi2026recommendation}.
Such utility–risk tensions may arise at different points in the memory lifecycle. We therefore organize memory operations into three functional stages: writing determines what information is stored, management transforms or reconciles stored information, and retrieval selects which memories are exposed as decision context \cite{wu2024longmemeval, latimer2025hindsight, lin2026survey}.
Each stage can improve benign memory utility, but it also governs a distinct pathway through which poisoned information is admitted, resurfaced, or suppressed.

\begin{figure}[t]
    \centering
    \includegraphics[width=0.9\linewidth]{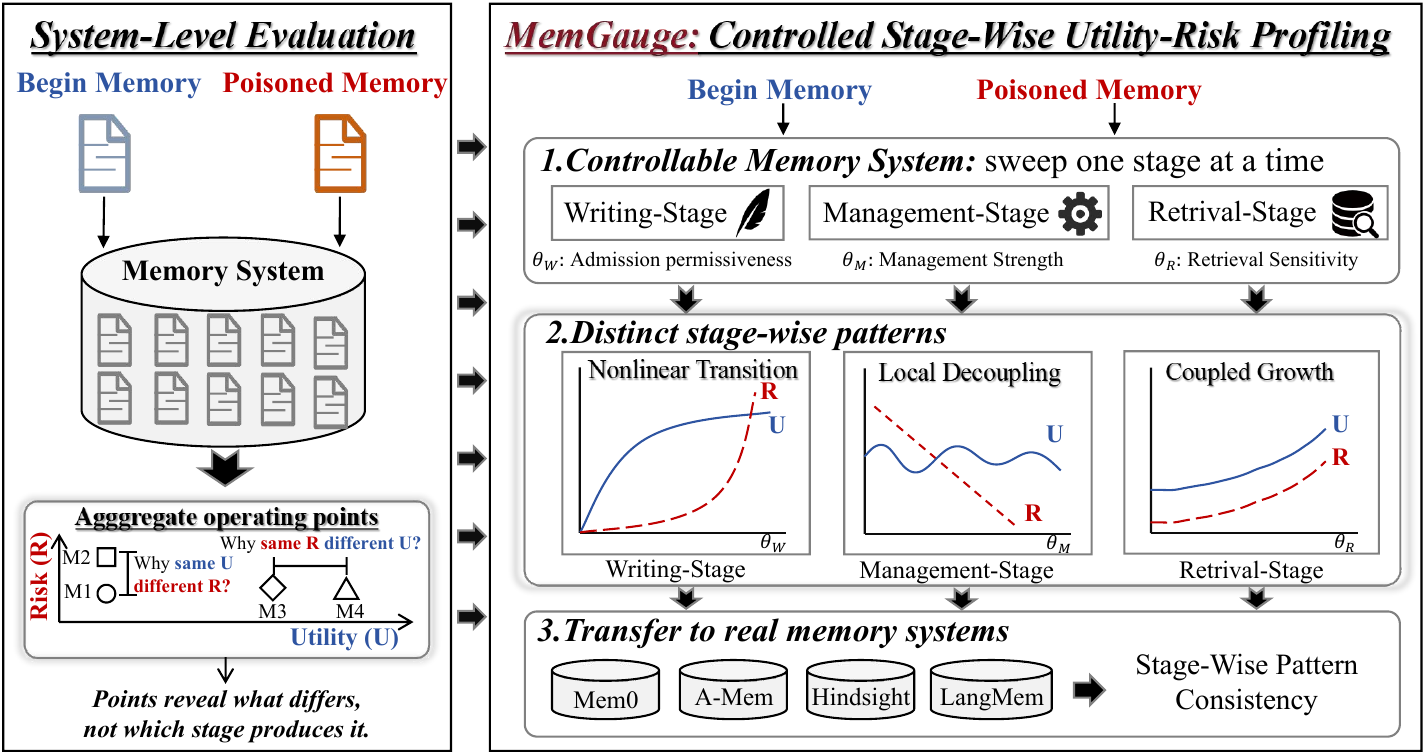}
    \caption{\textbf{Overview of stage-wise utility-risk evaluation.}
Unlike point estimates, \textsc{MemGauge} reveals stage-level patterns across memory systems.}
    \label{fig:indirect_poison}
    \vspace{-10pt}
\end{figure}

Existing evaluations report aggregate benign utility or poisoning risk for end-to-end memory systems, but rarely examine how these outcomes change with specific memory stages \cite{maharana2024evaluating,chhikara2025mem0,chen2024agentpoison,dong2026memory,piehl2026er,yang2026zombie}. Because each result combines the effects of writing, management, and retrieval, it identifies a system’s operating point but not how its utility–risk response changes within a stage. 
Changes at different stages may move utility and risk differently: one may increase both, whereas another may reduce risk with little utility loss. Without this stage-level view, a design intended to improve memory performance may yield only marginal utility gains while sharply increasing risk, while efforts to balance the two may rely on unguided trial and error across stages.
We therefore ask: \emph{How do controlled changes in writing admission, management policy, and retrieval exposure reshape memory utility and poisoning risk?}

Directly evaluating existing memory systems cannot readily answer this question. These systems tightly couple writing, management, and retrieval within end-to-end pipelines, expose only a small number of user-configurable interfaces, and differ in which controls they provide \cite{chhikara2025mem0,latimer2025hindsight}. Consequently, their measurements correspond to sparse and non-aligned operating points: a control available in one system may be absent or defined differently in another, making stage-wise utility–risk responses difficult to compare systematically.
To address this challenge, we introduce \textsc{MemGauge}, which varies one stage at a time in a controlled environment to construct stage-wise utility–risk response profiles, and then applies analogous stage-level measurements to existing systems to examine whether their observed operating points are consistent with these profiles.

Concretely, \textsc{MemGauge} defines three intervention axes corresponding to the memory lifecycle. At the writing stage, it divides each trace into atomic units, ranks them using the agent’s own LLM, and varies the ranked prefix passed to the writer to produce nested admission levels. At the management stage, it applies alternative conflict-handling policies to the same pre-management memory state. At the retrieval stage, it varies a shared exposure rate for helpful and risk-bearing records while keeping the retrieval budget fixed. At each operating point, paired clean and poisoned conditions hold the agent model, task, prompts, memory budget, and remaining memory operations fixed.

Using \textsc{MemGauge}, we conduct controlled evaluations of 11 LLMs across two long-term memory benchmarks. The resulting stage-wise utility–risk response profiles reveal three distinct patterns. \textit{\underline{(1) Writing stage}} exhibits a nonlinear transition: as admission coverage expands along each agent’s model-specific ranking, clean utility rises and gradually saturates, while attack success remains low at restrictive levels before increasing sharply. \textit{\underline{(2) Management stage}} exhibits policy-dependent local decoupling: changing the management policy can substantially shift attack success while leaving clean utility nearly unchanged, although the direction and magnitude of this shift depend on the policy. \textit{\underline{(3) Retrieval stage}} exhibits a coupled response: increasing the exposure of helpful and risk-bearing memories improves clean utility while simultaneously increasing attack success. These patterns show that utility and risk respond differently to changes at different stages of the memory lifecycle.

We further examine whether these patterns appear in four existing memory systems (e.g., Mem0 \cite{chhikara2025mem0}) using analogous stage-level measurements. For writing, we measure the fraction of trace units retained in memory; for management, we compare each system’s native mechanism when enabled and disabled on the same pre-management state; and for retrieval, we measure decision-relevant exposure across top-\(K\) budgets. Because these systems couple multiple stages, we interpret their operating points as diagnostic correspondence rather than controlled stage effects. The measurements align with the controlled profiles: greater information retention is associated with higher risk without reliable utility gains, management shifts attack success more than clean utility, and larger retrieval budgets raise both clean utility and attack success. These stage-level measurements help interpret aggregate end-to-end results.
 
Our contributions are threefold:
\begin{itemize}
    \item \textbf{Stage-wise utility–risk perspective.} We introduce a stage-wise utility-risk perspective spanning writing, management, and retrieval, complementing prior separate or aggregate system-level evaluations.
    
    \item \textbf{Controllable evaluation framework.} We develop \textsc{MemGauge}, which varies writing admission, compares management policies, and controls shared retrieval exposure under paired clean and poisoned conditions to construct stage-wise utility–risk response profiles.

    \item \textbf{Empirical pattern discovery.} Across 11 LLMs and two benchmarks, we observe a nonlinear transition in writing, policy-dependent local decoupling in management, and a coupled response in retrieval. Measurements on four existing memory systems exhibit patterns consistent with these profiles and help interpret their end-to-end behavior.

\end{itemize}

\section{Related Work}

Prior work studies agent memory from both utility and security perspectives. LoCoMo \cite{maharana2024evaluating}, LongMemEval \cite{wu2024longmemeval}, and MemBench \cite{tan2025membench} evaluate long-term memory across multi-session recall, temporal and cross-session reasoning, and knowledge updating. Systems such as Mem0 \cite{chhikara2025mem0} and A-MEM \cite{xu2026mem} further develop mechanisms for memory writing, retrieval, consolidation, and updating, typically evaluating the complete pipeline through task performance, retrieval quality, and efficiency.
Meanwhile, AgentPoison \cite{chen2024agentpoison}, MINJA \cite{dong2026memory}, MemoryGraft \cite{srivastava2025memorygraft}, Sleeper \cite{pulipaka2026hidden}, and other studies \cite{yang2026zombie,piehl2026er} show that poisoned memories or interactions can persistently influence downstream behavior, while MPBench \cite{zhou2025mpbench} systematizes memory-writing channels and structural vulnerabilities. However, existing work largely evaluates either end-to-end utility or attack effectiveness. Our work isolates writing, management, and retrieval to characterize their individual effects on the utility-risk tradeoff.

\section{\textsc{MemGauge} Framework}

We present \textsc{MemGauge}, a controlled framework for constructing stage-wise utility–risk response profiles of memory. It separately varies writing admission, management policy, and retrieval exposure while holding the other stages fixed, then applies analogous measurements to existing systems.
A summary of additional symbols, system prompts, and diagrams of methods and algorithms can be found in Section A of the Supplementary Materials.

\subsection{Problem Setting}

Let $i \in \{1,\ldots,N\}$ index an evaluation instance. Each instance contains a subsequent query $x_i$, its correct answer $y_i^\star$, an attacker-predefined target $y_i^{\mathrm{adv}} \neq y_i^\star$, and a pre-existing memory bank $B_i$ containing benign information irrelevant to the current query. We use $s \in \{c,p\}$ to denote the clean and poisoned conditions, while the superscripts $+$ and $-$ denote helpful and
risk-bearing information, respectively. The helpful trace $\tau_i^+$ contains information $e_i^+$ supporting $y_i^\star$, whereas the risk-bearing trace $\tau_i^-$ contains information $e_i^-$ supporting $y_i^{\mathrm{adv}}$. The clean condition contains only the helpful trace, while the poisoned condition
contains both traces. We consider interaction-based poisoning in which risk-bearing information enters memory through user- or tool-side interactions rather than direct memory edits.
The writer $W$ converts the two traces into memory records: \(\Delta M_i^+ = W(\tau_i^+)\) and \(\Delta M_i^- = W(\tau_i^-)\). The pre-management memory states are \(M_i^c = B_i \cup \Delta M_i^+\) and \(M_i^p = B_i \cup \Delta M_i^+ \cup \Delta M_i^-\). For either condition $s \in \{c,p\}$, management $G$ transforms the memory state \(\widetilde{M}_i^s = G(M_i^s)\), retrieval $R$ selects the context for the query \(X_i^s = R(x_i,\widetilde{M}_i^s)\), and the agent $F$ produces the final output: \(y_i^s = F(x_i,X_i^s)\).

\begin{figure*}[t]
    \centering
    \includegraphics[width=0.9\textwidth]{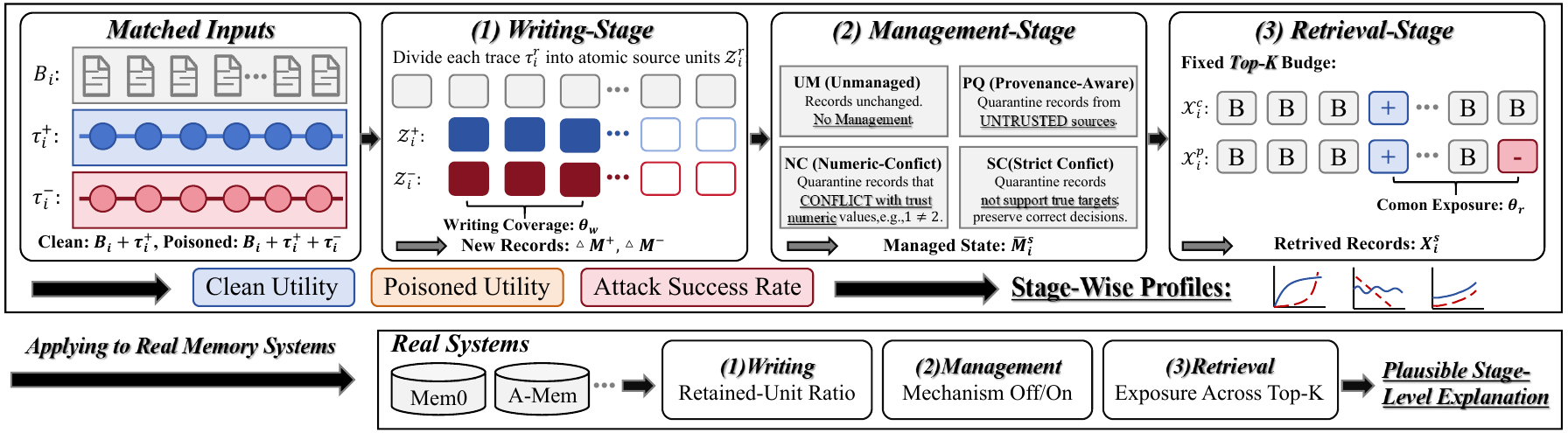}
    \caption{Overview of \textsc{MemGauge}. It varies one memory stage at a time under matched clean and poisoned conditions to construct utility-risk profiles and examine their diagnostic correspondence with existing memory systems.}
    \vspace{-10pt}
    \label{fig:framework}
\end{figure*}

At each operating point, \textsc{MemGauge} changes one stage-specific intervention while keeping the agent model, query, prompts, background memory, and remaining memory operations fixed. The paired outputs are used to compute clean utility, poisoned utility, and attack success rate, as defined in Section \ref{sec3.3}. Figure \ref{fig:framework}
summarizes this workflow: \textsc{MemGauge} first constructs controlled stage-wise response profiles and then compares them with analogous stage-level measurements from existing memory systems.

\subsection{Stage-Wise Interventions}
\textsc{MemGauge} constructs stage-wise response profiles within a controlled memory pipeline rather than by modifying existing systems. Within this pipeline, it defines one intervention axis for each stage
and varies one stage at a time under paired clean and poisoned conditions. We present the interventions in execution order: writing, management, and retrieval.

\mypara{(1) Writing: Controlling Information Admission.}
Although memory systems implement writing differently, they share a common purpose: extracting helpful information from interaction traces that may support future decisions and storing it as persistent records. \textsc{MemGauge} therefore represents writing in the controlled pipeline as admission followed by transformation. An admission controller selects the source units available at each operating point, while a fixed writer \(W\) converts the selected units into memory records. Holding \(W\) fixed ensures that differences across admission levels reflect changes in the available source information rather than changes in the writing mechanism itself.

Let $q \in \{+,-\}$ denote the helpful or risk-bearing information role. We divide each trace $\tau_i^q$ into a set of atomic source units $Z_i^q$. The same LLM used by the agent ranks the units within each
trace by their estimated usefulness for future decisions, producing an ordering $\pi_i^q$. The helpful and risk-bearing traces are ranked independently. At writing admission coverage $\theta_w \in [0,1]$, the controller supplies the corresponding ranked prefix to $W$:
\begin{equation}
b_i^q(\theta_w)
=
\left\lfloor \theta_w |Z_i^q| \right\rfloor,
A_i^q(\theta_w)
=
\left\{
\pi_i^q(1),\ldots,
\pi_i^q\bigl(b_i^q(\theta_w)\bigr)
\right\}.
\end{equation}
where $A_i^q(0)=\emptyset$. Because these admitted sets are nested, increasing $\theta_w$ only adds lower-ranked units from the same trace.
The writer converts the admitted units into memory records:
\(
\Delta M_i^q(\theta_w)=\bigcup_{z\in A_i^q(\theta_w)} W(z).
\)
The resulting clean and poisoned memory states are
\begin{equation}
M_i^c(\theta_w)
=
B_i \cup \Delta M_i^+(\theta_w),
\end{equation}
\begin{equation}
M_i^p(\theta_w)
=
B_i \cup \Delta M_i^+(\theta_w)
\cup \Delta M_i^-(\theta_w).
\end{equation}

Each source unit is processed by the writer once, and its resulting records are reused at every coverage level containing that unit. Thus, differences across $\theta_w$ arise from the admitted source units rather than repeated writer generation.

\mypara{(2) Management: Controlling Memory-State Transformation.}
Although memory systems implement management differently, these mechanisms share a common purpose: transforming stored information before retrieval by filtering, reconciling, or updating memory records. Unlike writing admission, these operations do not form a natural continuous axis. MemGauge therefore represents management using discrete policies applied to the same pre-management memory state.
For each condition $s \in \{c,p\}$, a policy $g$ transforms the memory state as
\begin{equation}
\widetilde{M}_i^s(g)
=
G_g(M_i^s),
g \in
\mathcal{G}
=
\{\mathrm{UM},\mathrm{PQ},\mathrm{NC},\mathrm{SC}\}.
\end{equation}
Here, quarantining a record means excluding it from subsequent retrieval while leaving the remaining memory state unchanged.
\emph{Unmanaged Memory} (UM) leaves all records unchanged and serves as the no-management reference. \emph{Provenance-Aware Quarantine} (PQ) quarantines records written from sources marked as untrusted. \emph{Numeric-Conflict Resolution} (NC) quarantines a newly written record when its numerical content conflicts with an existing trusted record. \textit{Strict Conflict Resolution (SC)} retains records consistent with verified interaction outcomes, such as explicit user confirmations or tool-execution results, and quarantines those that are not. It represents an idealized setting in which such feedback is
complete and reliable, and serves as an upper bound.

Within each condition, all policies receive the same pre-management state $M_i^s$, while the subsequent retrieval configuration and agent remain fixed. The policies therefore represent separate management operating points rather than ordered levels of management capability.

\mypara{(3) Retrieval: Controlling Decision-Relevant Exposure.}
Retrieval determines which records from the memory are exposed to the agent for the current query. Although retrieval mechanisms differ in how they rank records, their decision-relevant outcome is whether helpful or risk-bearing information enters the retrieved context under a fixed retrieval budget. \textsc{MemGauge} therefore intervenes directly on record exposure rather than modifying any particular ranking algorithm.

For each instance, the retriever returns a context containing exactly $K$ records. At retrieval exposure probability $\theta_r\in[0,1]$, we sample
\begin{equation}
a_i^{+}\sim\operatorname{Bernoulli}(\theta_r),
a_i^{-}\sim\operatorname{Bernoulli}(\theta_r),
\end{equation}
where $a_i^{+}$ and $a_i^{-}$ indicate whether the helpful record $e_i^{+}$ and the risk-bearing record $e_i^{-}$ are included in the retrieved context, respectively. In particular,
\begin{equation}
\Pr(a_i^{+}=1)
=
\Pr(a_i^{-}=1)
=
\theta_r.
\end{equation}
The clean and poisoned conditions share the same helpful-record exposure decision $a_i^{+}$, while the risk-bearing record is available only in the poisoned condition.
The retrieved contexts are constructed as
\begin{equation}
X_i^{c}(\theta_r)
=
a_i^{+}\{e_i^{+}\}
\cup
\operatorname{Fill}
\left(
B_i,
K-a_i^{+}
\right),
\end{equation}
\begin{equation}
X_i^{p}(\theta_r)
=
a_i^{+}\{e_i^{+}\}
\cup
a_i^{-}\{e_i^{-}\}
\cup
\operatorname{Fill}
\left(
B_i,
K-a_i^{+}-a_i^{-}
\right).
\end{equation}
Here, $a\{e\}$ denotes $\{e\}$ when $a=1$ and the empty set otherwise. The function $\operatorname{Fill}(B_i,k)$ selects $k$ task-irrelevant records from $B_i$ according to a fixed ordering. When the risk-bearing record is exposed, it replaces an irrelevant background record rather than the helpful record.
The two conditions use the same helpful-record exposure decision, background-record ordering, and record positions, and both always contain exactly $K$ records. We use balanced sampling schedules so that the empirical exposure frequency at each operating point matches $\theta_r$.

\subsection{Constructing Stage-Wise Utility-Risk Profiles}
\label{sec3.3}

At each operating point, we evaluate the same instances under matched clean and poisoned conditions while holding the model, queries, prompts, decoding settings, and remaining memory operations fixed. From the paired outputs, we compute clean utility, poisoned utility, and attack success rate (ASR). Clean utility measures correct decisions under the clean condition, poisoned utility measures correct decisions after risk-bearing information is introduced, and ASR measures decisions redirected to the attacker-predefined target. Poisoned utility and ASR are not complementary because an output may match neither target.
Varying $\theta_w$ or $\theta_r$ produces the writing- and retrieval-stage response profiles, respectively. Because management policies do not form an ordered scale, each policy is represented as a separate utility-risk point. These profiles characterize responses along stage-specific intervention axes and do not place writing, management, and retrieval on a shared capability scale.

\subsection{Applying \textsc{MemGauge} to Memory Systems}
Existing memory systems couple writing, management, and retrieval and do not expose the same intervention parameters as \textsc{MemGauge}. We therefore do not directly impose $\theta_w$, $\theta_r$, or the controlled management policies on these systems. Instead, we construct analogous stage-level measurements using the same helpful and risk-bearing traces, background memory, and subsequent queries.
For writing, we measure the proportion of source units whose information is retained in the resulting memory records. For management, we compare the system's native mechanism when enabled and disabled on the same pre-management state. For retrieval, we vary the top-$K$ budget and measure the proportion of instances in which the risk-bearing record appears in the retrieved context. We then examine whether the resulting utility and ASR patterns qualitatively correspond to the controlled profiles. Because these measurements come from coupled system pipelines, they are treated as diagnostic associations rather than controlled stage effects.

\section{Experiments}

\subsection{Experimental Setup}

\mypara{Agent and LLMs.}
We implement a common LangGraph agent (38.4K GitHub stars) with 11 LLMs, including GPT-5.4, GPT-5.4-Mini, Gemini 3.1 Flash, Gemini 2.5 Flash, Grok 4.3 High, Grok 4.2 Fast, DeepSeek-V4-Flash, GLM-5.1, Qwen 3.5 397B, MiniMax-M3, and Doubao Seed 2.0 Pro. For each configuration, the same LLM serves as the agent's reasoning model and the writing-stage ranker.

\mypara{Datasets and trace construction.}
We construct evaluation instances from LongMemEval \cite{wu2024longmemeval} and LoCoMo \cite{maharana2024evaluating}, selecting 100 test questions from each dataset and identifying the information required to answer them. This information is introduced through a benign agent interaction, producing the helpful trace $\tau_i^{+}$. The background memory bank $B_i$ is constructed from analogous interactions generated using information from the corresponding dataset that is unrelated to the subsequent query.
For the poisoned condition, MINJA \cite{dong2026memory} and Sleeper \cite{pulipaka2026hidden} generate user-side and tool-side risk-bearing interactions, respectively, producing $\tau_i^{-}$. The attacker can influence the corresponding interaction content but cannot directly edit the memory bank. Helpful and risk-bearing traces are generated in separate sessions, and the test query is issued in a subsequent session. In the controlled experiments, traces are processed by the fixed writer $W$; in the existing-system evaluation, they are processed by each system's native writing mechanism.

\mypara{Existing memory systems.}
We consider four existing memory system: Mem0 (61.8K GitHub stars) \cite{chhikara2025mem0}, Hindsight (18.8K GitHub stars) \cite{latimer2025hindsight}, LangMem (1.6K GitHub stars) \cite{langchain2025langmem}, and A-Mem (1.1K GitHub stars) \cite{xu2026mem}.

\mypara{Metrics.}
We report clean utility, poisoned utility, and ASR as defined in Section~\ref{sec3.3}. AgentEvals \cite{langchain_agentevals} judges whether each trajectory supports the correct answer or the attacker-predefined target. Its agreement with human annotations is evaluated in Section~\ref{sec:human_eval}.

\mypara{Parameters.}
Unless otherwise specified, we set the retrieval budget to $K=5$, and the background memory size $|B_i|=100$. Matched clean and poisoned evaluations use the same configuration (e.g., model, system prompt, query). When evaluating one stage, the other stages are fixed at \(\theta_w=0.6\), UM, and \(\theta_r=0.8\), respectively.

\mypara{Research questions.}
Our experiments address the following research questions.
\textbf{\textit{\underline{RQ1}}}: What utility--risk profiles emerge under controlled writing, management, and retrieval interventions?
\textbf{\textit{\underline{RQ2}}}: How stable are the observed stage-wise profiles across evaluation
conditions?
\textbf{\textit{\underline{RQ3}}}: Do analogous stage-level measurements on existing memory systems exhibit qualitative correspondence with the controlled profiles?

\subsection{RQ1: Stage-Wise Utility-Risk Profiles}

\begin{figure*}[t]
    \centering
    \includegraphics[width=\textwidth]{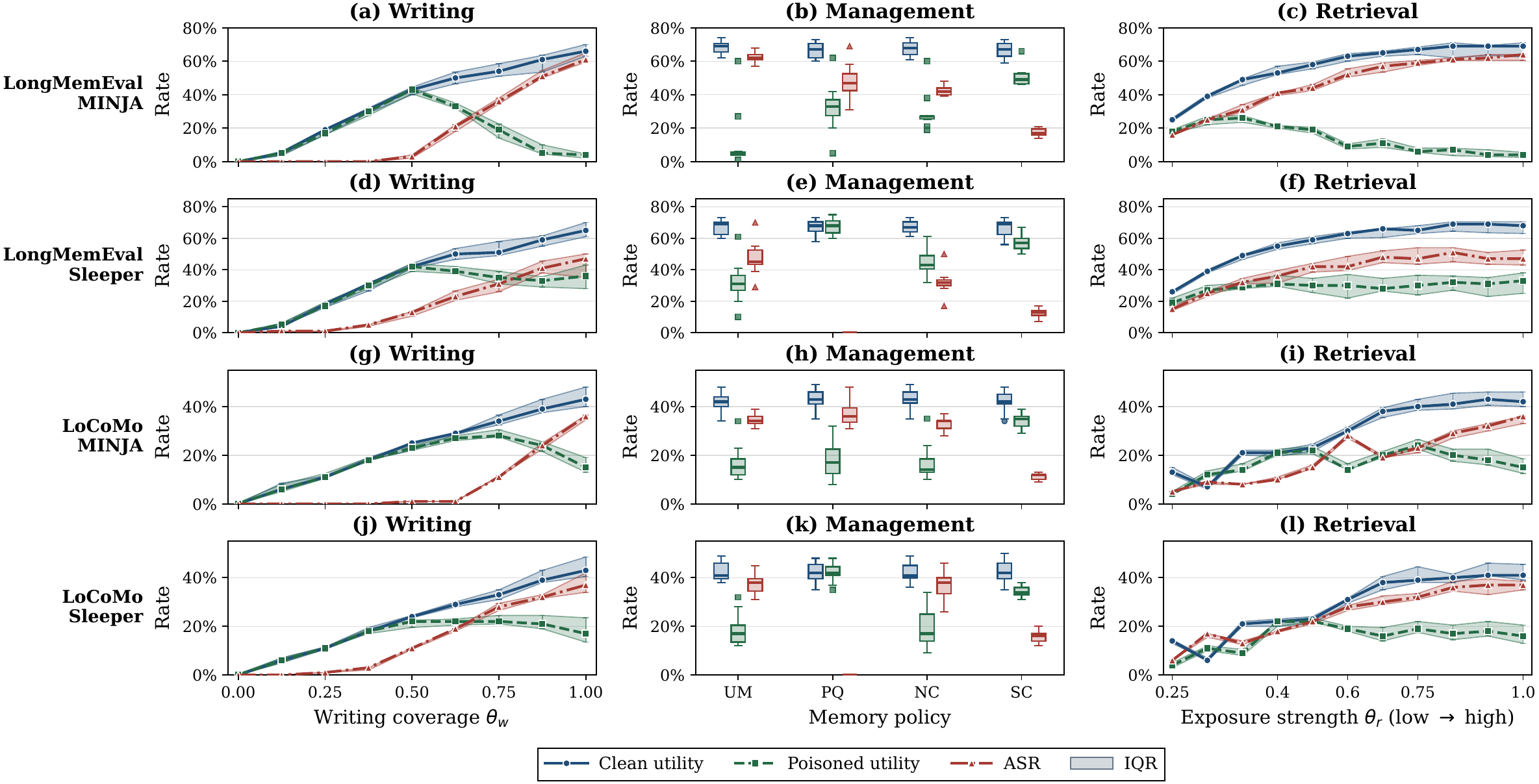}
    \caption{Stage-wise utility-risk profiles across 11 LLMs. Blue, green, and
red denote clean utility, poisoned utility, and ASR; lines show the median,
while shaded bands and box plots indicate the IQR.}
    \label{fig:stage_profiles}
\end{figure*}

Figure ~\ref{fig:stage_profiles} presents the controlled stage-wise profiles across 11 LLMs. Each column varies one memory stage while holding the remaining operations fixed, and each row represents a dataset--attack combination. Lines report the median across models, with shaded regions and box plots showing the interquartile range. Per-model results are in Supplementary Material Section B.1.

\mypara{Writing-Stage Profile.}
At low coverage ($\theta_w<0.3$), clean and poisoned utility increase almost synchronously to approximately $15\%$-$20\%$, while ASR remains near zero. Between $\theta_w=0.3$ and $0.5$, both utility measures continue to rise, reaching roughly $20\%$-$42\%$, but ASR generally remains below $10\%$. Beyond this range, the profiles diverge sharply: clean utility continues increasing, whereas poisoned utility reverses its trend and declines from its peak to $3\%$-$34\%$ at full coverage. Meanwhile, ASR rises rapidly to $37\%$-$64\%$. This delayed but sharp divergence reveals a threshold-like nonlinearity: additional risk-bearing information has little downstream effect at low coverage but becomes increasingly influential once sufficient content is retained.

\mypara{Management-Stage Profile.}
Management can effectively mitigate memory risks, but its effectiveness differs substantially across attack types. Across the evaluated policies, clean utility varies by only $3\%$-$5\%$, whereas ASR changes by up to $62\%$, indicating that management can locally decouple risk from clean performance. However, the extent of this decoupling is mechanism-specific. PQ reduces Sleeper ASR to nearly zero with little utility change but provides limited protection against MINJA, whereas SC achieves more consistent risk reductions across both attacks.

\mypara{Retrieval-Stage Profile.}
Retrieval exhibits a coupled utility-risk response. As retrieval exposure increases, clean utility rises from approximately $10\%$--$25\%$ to $43\%$-$70\%$, while ASR simultaneously increases from about $5\%$-$18\%$ to $34\%$-$64\%$. At high exposure, clean utility begins to saturate, whereas ASR remains high and poisoned utility falls to approximately $4\%$-$31\%$. This coupling is expected because retrieval ranks records by their relevance to the query rather than determining whether their content is benign or risk-bearing; a sufficiently relevant risk-bearing record can therefore receive exposure comparable to a helpful record.

\begin{tcolorbox}[
    colback=black!2,
    colframe=black!35,
    boxrule=0.5pt,
    arc=2pt,
    left=7pt,
    right=7pt,
    top=5pt,
    bottom=5pt
]
\textbf{Key Finding 1.}
Writing exhibits a \emph{threshold-like risk transition}, management enables \emph{mechanism-dependent local decoupling}, and retrieval \emph{couples utility with risk}. Effective control therefore requires calibrating writing coverage to avoid the high-risk region, adopting mechanism-matched management policies, and constraining retrieval exposure rather than expecting relevance-based ranking to distinguish helpful from risk-bearing records.
\end{tcolorbox}

\subsection{RQ2: Profile Stability}

We examine how selected workload and implementation factors affect the stage-wise profiles. Unless otherwise specified, all experiments in this section use DeepSeek-V4-Flash, LongMemEval, and MINJA.

\begin{figure}[h]
    \centering
    \includegraphics[width=1\linewidth]{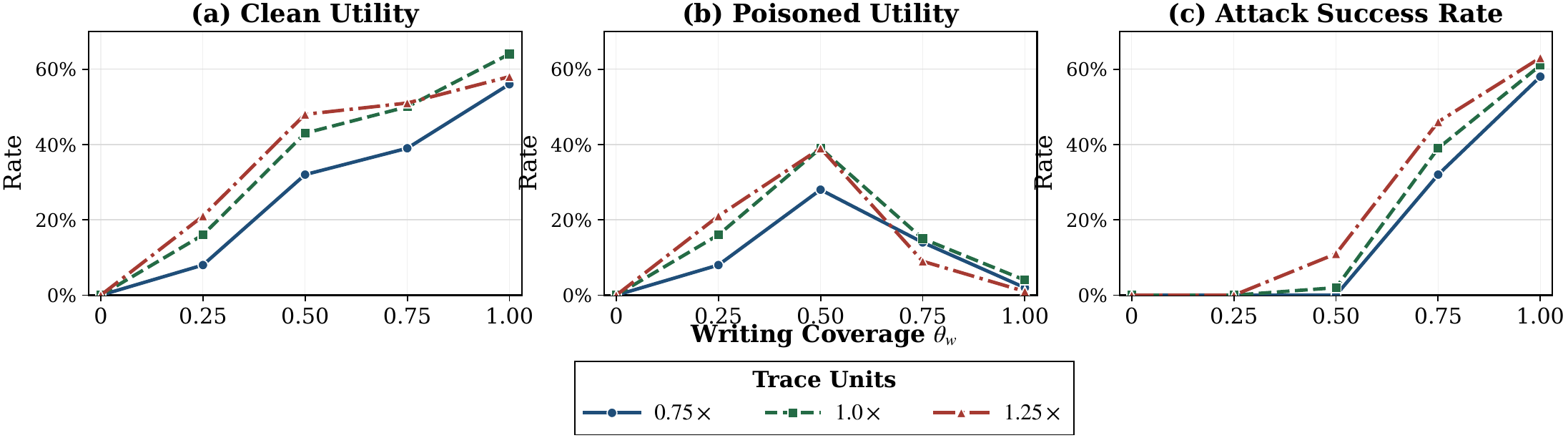}
    \caption{Writing-stage profiles across different trace loads.}
    \label{fig:trace_stability}
    \vspace{-10pt}
\end{figure}

\mypara{Trace-Load stability.}
As shown in Figure \ref{fig:trace_stability}, we vary the number of source units in each trace to $0.75\times$, $1.0\times$, and $1.25\times$ the default size while preserving the core helpful or risk-bearing information. Across all three settings, ASR remains near zero at low writing coverage and increases more sharply at higher coverage. At full coverage, ASR reaches $58\%$-$63\%$, while poisoned utility falls to $1\%$-$4\%$. Larger trace loads shift the rise in risk toward lower coverage levels but preserve the overall writing-stage response.

\begin{figure}[h]
    \centering
    \includegraphics[width=1\linewidth]{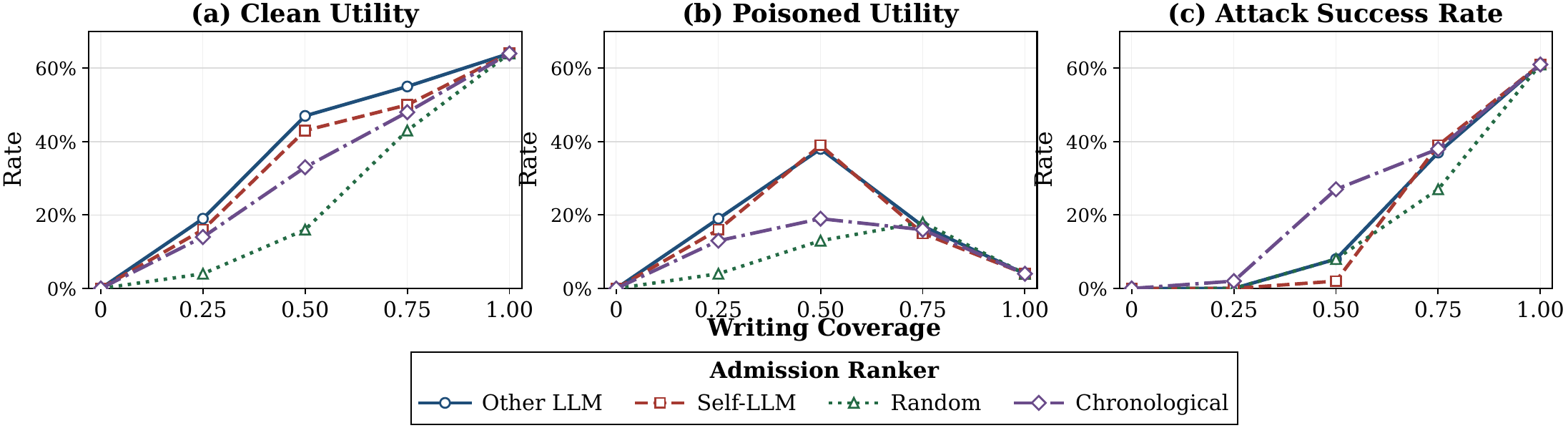}
    \caption{Writing-stage profiles under different unit rankers.}
    \label{fig:ranker_stability}
    \vspace{-10pt}
\end{figure}

\mypara{Ranker stability.}
As shown in Figure \ref{fig:ranker_stability}, we compare four source-unit orderings: the agent's own LLM, an external LLM, random ordering, and chronological ordering. At $\theta_w=0.5$, the self-LLM and external-LLM rankers achieve $43\%$--$47\%$ clean utility with $2\%$-$8\%$ ASR, whereas random ordering reaches only $16\%$ clean utility. Chronological ordering achieves $33\%$ clean utility but raises ASR to $27\%$. All orderings converge at $\theta_w=1$ because full coverage includes the same source units by construction. Thus, ranker choice shifts the utility-risk trajectory and the onset of risk without changing the full-coverage endpoint.

\begin{figure}[h]
    \centering
    \includegraphics[width=1\linewidth]{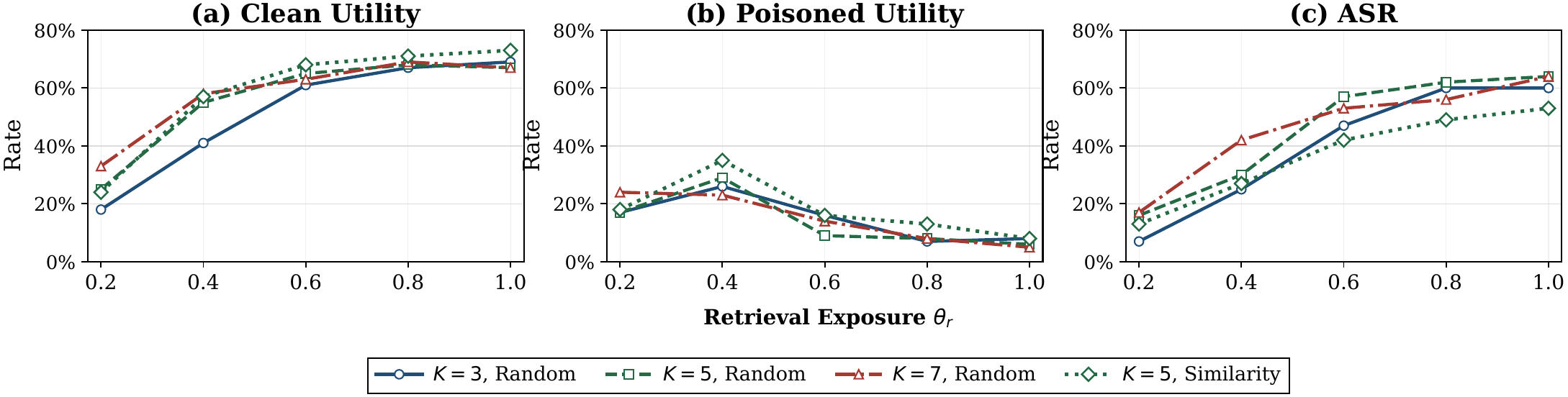}
    \caption{Retrieval-stage profiles across retrieval budgets and distractor-filling strategies.}
    \label{fig:k_stability}
    \vspace{-10pt}
\end{figure}

\mypara{Retrieval-budget stability.}
Because \(K\) directly controls the amount of memory returned to the agent, we evaluate its effect primarily at the retrieval stage. As shown in Figure~\ref{fig:k_stability}, across all budgets, increasing $\theta_r$ raises both clean utility and ASR under matched helpful and risk-bearing exposure. The differences are most visible at low and intermediate exposure levels, while the profiles converge at high exposure. Thus, $K$ shifts the utility-risk operating points but does not remove the coupled response induced by matched exposure.

\begin{figure}[h]
    \centering
    \includegraphics[width=1\linewidth]{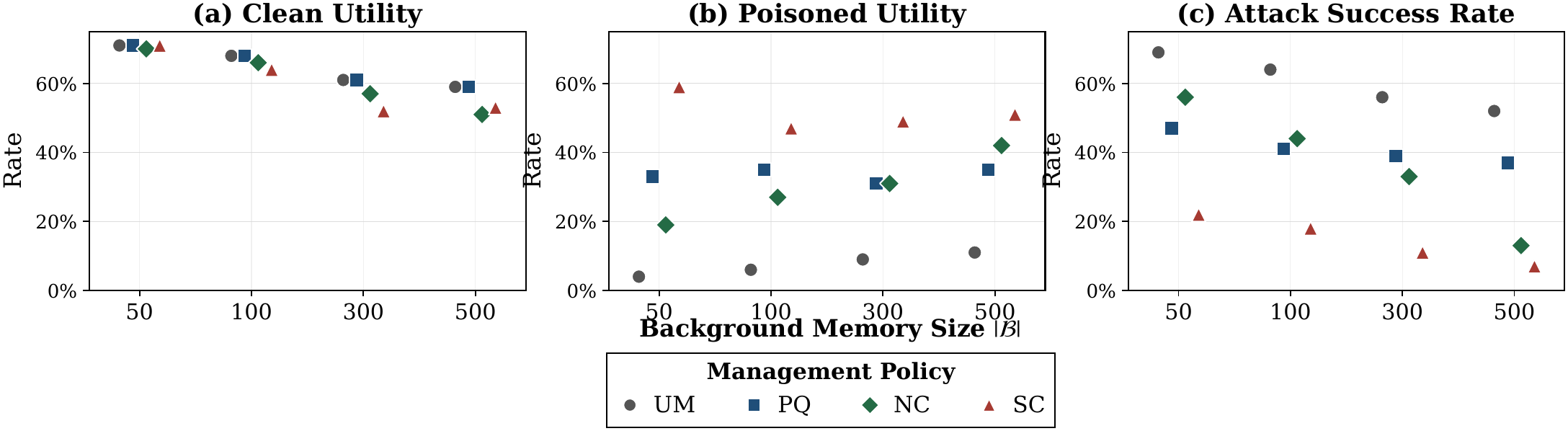}
    \caption{Management-stage profiles across memory sizes.}
    \label{fig:b_stability}
    \vspace{-10pt}
\end{figure}

\mypara{Distractor-fill stability.}
Under a fixed retrieval budget of $K=5$, we replace random background filling with similarity-based filling. After excluding records that support either the correct answer or the attacker-predefined target, we use the most query-similar remaining records to fill unused retrieval slots. 
As shown in Figure \ref{fig:k_stability}, the composition of background memories shifts the utility-risk
operating point but does not remove the coupled retrieval-stage response.

\mypara{Asymmetric-exposure stability.}
We fix the helpful-record exposure probability at $\theta_r^{+}=0.8$ and vary the risk-bearing exposure probability $\theta_r^{-}$ from $0.2$ to $1.0$. ASR increases steadily from $0.13$ to $0.63$, indicating that greater risk-bearing exposure consistently amplifies attack risk even when helpful-record exposure remains fixed. Detailed results are provided in Section B.2 of the supplementary material.

\mypara{Memory-scale stability.}
Changing the memory-bank size primarily affects management and retrieval. Increasing \(|B|\) intensifies retrieval competition, resembling a tighter effective budget. Since retrieval sensitivity is examined above, we focus here on management and vary \(|B|\) from \(50\) to \(500\).
As shown in Figure \ref{fig:b_stability}, when the memory bank grows, retrieval competition reduces both clean utility and ASR under UM. Nevertheless, management continues to produce distinct utility--risk shifts. At $|B_i|=500$, PQ reduces ASR from $52\%$ to $37\%$ without changing clean utility, while NC and SC achieve larger ASR reductions with clean-utility decreases of $8$ and $6$ percentage points, respectively.

\begin{tcolorbox}[
    colback=black!2,
    colframe=black!35,
    boxrule=0.5pt,
    arc=2pt,
    left=7pt,
    right=7pt,
    top=5pt,
    bottom=5pt
]
\textbf{Key Finding 2.}
Within the evaluated model-dataset-attack setting, the selected trace, ranking, retrieval, distractor, and memory-scale factors shift the onset and magnitude of the responses while preserving broadly similar qualitative stage-wise patterns.
\end{tcolorbox}

\subsection{RQ3: Correspondence in Existing Systems}
We examine whether analogous stage-level measurements on Mem0, Hindsight, LangMem, and A-MEM exhibit qualitative correspondence with the controlled profiles. Table~\ref{tab:real_system_wm} reports the writing and management results. WR denotes the proportion of source units retained by the system's writer, $U_0$ denotes utility without memory, and $M_0$ and $M_1$ denote native management disabled and enabled, respectively. Unless otherwise specified, we use $|B_i|=100$ and $K=5$.

\begin{table}[h]
\centering
\caption{Writing and management results on real-world memory systems. WR is
reported as mean \(\pm\) standard deviation over 500 writing trials.}
\label{tab:real_system_wm}
\footnotesize
\setlength{\tabcolsep}{2.5pt}
\begin{tabular}{@{}lcccccc@{}}
\toprule
System & WR (\%) & \(U0\) & Mode & Clean U & Poisoned U & ASR \\
\midrule
\multirow{2}{*}{Mem0}
 & \multirow{2}{*}{\(36.5\pm6.1\)}
 & \multirow{2}{*}{0.008}
 & \(M0\) & 0.533 & 0.213 & 0.326 \\
 & & & \(M1\) & 0.541 & 0.343 & 0.176 \\
\addlinespace[2pt]

\multirow{2}{*}{Hindsight}
 & \multirow{2}{*}{\(69.4\pm5.4\)}
 & \multirow{2}{*}{0.015}
 & \(M0\) & 0.563 & 0.165 & 0.475 \\
 & & & \(M1\) & 0.524 & 0.240 & 0.293 \\
\addlinespace[2pt]

\multirow{2}{*}{LangMem}
 & \multirow{2}{*}{\(74.5\pm8.7\)}
 & \multirow{2}{*}{0.011}
 & \(M0\) & 0.528 & 0.158 & 0.473 \\
 & & & \(M1\) & 0.523 & 0.325 & 0.225 \\
\addlinespace[2pt]

\multirow{2}{*}{A-MEM}
 & \multirow{2}{*}{\(55.7\pm7.7\)}
 & \multirow{2}{*}{0.008}
 & \(M0\) & 0.483 & 0.130 & 0.403 \\
 & & & \(M1\) & 0.460 & 0.218 & 0.297 \\
\bottomrule
\end{tabular}
\end{table}

\mypara{Writing-stage correspondence.}
Writing exposure varies substantially across the four systems, from \(36.5\%\) in Mem0 to \(74.5\%\) in LangMem. Meanwhile, \(U0\) remains only \(0.8\%\)-\(1.5\%\), confirming that the evaluated tasks largely depend on information supplied through memory. Under \(M0\), higher writing exposure broadly coincides with greater risk. Mem0 has both the lowest WR and ASR (\(36.5\%\) and \(32.6\%\)), whereas Hindsight and LangMem exhibit high WR (\(69.4\%\)-\(74.5\%\)) and ASR (\(47.3\%\)-\(47.5\%\)). However, clean utility does not increase correspondingly: Mem0 achieves \(53.3\%\) clean utility despite its substantially lower WR, comparable to LangMem at \(52.8\%\). These results suggest that greater writing exposure expands risk exposure without guaranteeing additional clean utility.

\mypara{Management-stage correspondence.}
Enabling native management reduces observed ASR in all four systems by $10.6$-$24.8$ percentage points and increases poisoned utility by $7.5$-$16.7$ percentage points. The corresponding clean-utility changes range from a $0.8$-percentage-point increase to a $3.9$-percentage-point decrease. LangMem shows the largest ASR reduction, from $47.3\%$ to $22.5\%$, with only a $0.5$-percentage-point decrease in clean utility. These results qualitatively correspond to the policy-dependent utility--risk shifts observed in the controlled management profiles.

\begin{table}[h]
\centering
\caption{Retrieval results under different retrieval budgets.}
\label{tab:real_system_retrieval}
\footnotesize
\setlength{\tabcolsep}{3.2pt}
\renewcommand{\arraystretch}{0.92}
\begin{tabular}{@{}llcccc@{}}
\toprule
System & Metric & \(K=1\) & \(K=3\) & \(K=5\) & \(K=7\) \\
\midrule
\multirow{4}{*}{Mem0}
 & RSR        & 0.510 & 0.790 & 0.920 & 0.970 \\
 & Clean U    & 0.287 & 0.417 & 0.533 & 0.586 \\
 & Poisoned U & 0.246 & 0.245 & 0.213 & 0.194 \\
 & ASR        & 0.047 & 0.249 & 0.326 & 0.361 \\
\addlinespace[2pt]
\midrule
\multirow{4}{*}{Hindsight}
 & RSR        & 0.460 & 0.730 & 0.880 & 0.930 \\
 & Clean U    & 0.293 & 0.472 & 0.563 & 0.642 \\
 & Poisoned U & 0.241 & 0.224 & 0.165 & 0.134 \\
 & ASR        & 0.086 & 0.359 & 0.475 & 0.537 \\
\addlinespace[2pt]
\midrule
\multirow{4}{*}{LangMem}
 & RSR        & 0.470 & 0.720 & 0.840 & 0.910 \\
 & Clean U    & 0.287 & 0.463 & 0.528 & 0.613 \\
 & Poisoned U & 0.225 & 0.193 & 0.158 & 0.147 \\
 & ASR        & 0.094 & 0.328 & 0.473 & 0.526 \\
\addlinespace[2pt]
\midrule
\multirow{4}{*}{A-MEM}
 & RSR        & 0.420 & 0.650 & 0.790 & 0.860 \\
 & Clean U    & 0.246 & 0.395 & 0.483 & 0.527 \\
 & Poisoned U & 0.213 & 0.176 & 0.130 & 0.095 \\
 & ASR        & 0.122 & 0.313 & 0.403 & 0.478 \\
\bottomrule
\end{tabular}
\end{table}

To evaluate retrieval, we vary the retrieval budget \(K\) as a proxy for increasing memory exposure; all experiments are conducted under \(M0\) with \(|{B}_i|=100\), and RSR denotes the proportion of instances in which the target record (helpful or risk) appears in the top-\(K\) retrieved context.

\mypara{Retrieval-stage correspondence.}
Increasing the retrieval budget produces a consistent coupled utility-risk response across all four systems. As \(K\) increases from \(1\) to \(7\), RSR rises from \(42\%\)-\(51\%\) to \(86\%\)-\(97\%\), while clean utility increases from \(24.6\%\)-\(29.3\%\) to \(52.7\%\)-\(64.2\%\). This utility gain is accompanied by a substantial increase in ASR, from \(4.7\%\)-\(12.2\%\) to \(36.1\%\)-\(53.7\%\), and a decrease in poisoned utility from \(21.3\%\)-\(24.6\%\) to \(9.5\%\)-\(19.4\%\). The strength of this response varies across systems: Hindsight gains \(34.9\%\) in clean utility but also increases ASR by \(45.1\%\), whereas Mem0 shows a smaller ASR increase of \(31.4\%\) alongside a \(29.9\%\) utility gain. Nevertheless, the direction is consistent across implementations. This behavior is expected because retrieval ranks records by query relevance rather than distinguishing benign from risk-bearing content; increasing \(K\) therefore raises the exposure of both helpful and well-matched risk-bearing records.

\begin{tcolorbox}[
    colback=black!2,
    colframe=black!35,
    boxrule=0.5pt,
    arc=2pt,
    left=7pt,
    right=7pt,
    top=5pt,
    bottom=5pt
]
\textbf{Key Finding 3.}
The four existing systems exhibit qualitative correspondence with the controlled profiles: writing exposure raises risk without reliable utility gains, native management shifts the observed operating points toward lower ASR, and larger retrieval budgets are accompanied by increases in both clean utility and ASR. These results are diagnostic associations rather than controlled stage effects.
\end{tcolorbox}

\subsection{Human Evaluation}
\label{sec:human_eval}
We validate AgentEval using 200 clean and 200 poisoned trajectories annotated independently by five human experts. Annotators judge whether each output is correct or, for poisoned trajectories, matches the attacker-predefined target; the final label is determined by majority vote. AgentEval achieves \(91\%\) agreement with the human labels, while inter-annotator agreement is \(\kappa=0.78\), supporting its use for evaluation.

\section{Conclusion}

This paper introduced \textsc{MemGauge}, a paired, stage-wise framework for evaluating utility and risk in long-term memory systems. Across the evaluated settings, writing shows a threshold-like risk transition, management enables mechanism-dependent local decoupling, and retrieval couples utility gains with increased risk. These profiles persist across configurations and four existing memory systems, although their onset and magnitude vary. The results show that end-to-end metrics alone are insufficient and motivate stage-specific controls.
\bibliographystyle{plain}
\bibliography{reference}

\end{document}